
%
%
\magnification=\magstep1
\openup 1\jot
\def\xb{{\bf x}}
\font\secbf=cmbx12
\font\subsecbf=cmbx10

\outer\def\beginsec#1\par{\vskip0pt plus.3\vsize\penalty-250
 \vskip0pt plus-.3\vsize\bigskip\vskip\parskip
 \message{#1}\leftline{\secbf#1}\nobreak\smallskip\noindent}

\outer\def\beginsubsec#1\par{\vskip0pt plus.1\vsize\penalty-100
 \vskip0pt plus-.1\vsize\medskip\vskip\parskip
 \message{#1}\leftline{\subsecbf#1}\nobreak\smallskip\noindent}

\centerline{{\bf THE LEAST ACTION PRINCIPLE}}
\centerline{{\bf AND}}
\centerline{{\bf THE SPIN OF GALAXIES IN THE LOCAL GROUP}}
\vskip 0.5 truein
\centerline{ A.M. Dunn\footnote{$^ 1$}{Email: amd@mail.ast.cam.ac.uk}}
\centerline{Institute of Astronomy}
\centerline{University of Cambridge, Madingley Road}
\centerline{Cambridge, CB3 0HA, UK}
\vskip 0.5 truein
\centerline{ R. Laflamme\footnote{$^ 2$}{Email: laf@tdo-serv.lanl.gov}}
\centerline{Theoretical Astrophysics, T-6, MSB288}
\centerline{Los Alamos National Laboratory}
\centerline{Los Alamos, NM87545, USA}
\vskip 1 truein
\centerline{\bf Abstract}
\bigskip
Using  Peebles' least action principle,  we determine trajectories for the
galaxies in the Local Group and the more massive galaxies in the Local
Neighbourhood. We deduce the resulting angular momentum for the whole of the
Local Group and study the tidal force
acting on the Local Group and its galaxies. Although Andromeda and the
Milky Way dominate the tidal force acting on each other during the present
epoch,
we show that there is a transition time at $z\approx 1$ before which the
tidal force is dominated by galaxies outside the Local Group in each case. This
shows that the Local Group can not be considered as an isolated system as far
as the tidal forces are concerned.  We integrate the tidal torques acting on
the Milky Way and Andromeda and derive their spin angular momenta, obtaining
results which are comparable with observation.

\vfill \eject

\beginsec  1. Introduction.

One of the problems facing any theory of galaxy formation is to explain how
proto-galaxies can acquire spin angular momentum as they emerge from a
non-rotating  homogeneous distribution of gas. One of the early solutions was
presented by Hoyle(1949), who suggested that the spin of a galaxy might
result from tidal interactions with its neighbours. Hoyle's analysis
was set against the background of a steady state universe, but the idea
extends to an expanding universe scenario.
It has been shown that tidal interactions can produce galaxies with
the rotations of the right order of magnitude (Peebles 1969, White 1984,
Barnes \& Efstathiou 1987), although results beyond this remain elusive.

A major problem is that in order to calculate the torque on a given galaxy,
its history and the histories of its neighbours must be known. Although
we have relatively accurate positions for galaxies in the vicinity of
the Local Group, we have only the radial component of their velocities.
It is therefore very hard to trace their orbits back in time. Moreover, if
this is attempted, the uncertainty in the velocities increases without bound
due to unstable solutions of the equations of motion. A small error in the
velocity now would translate into an enormous difference in position in the
early universe. Thus it seems impossible to trace back galaxies using only
the observational data.

Peebles (1989,1990) proposed a method he calls the least action
principle, which he used to find complete trajectories for Local Group
galaxies.
The idea is to assume that galaxies growing out of small density perturbations
in the early universe will have negligible peculiar velocities with respect
{}to the Hubble flow. Using this as one boundary condition and the present
positions of the galaxies as the other, trial orbits are iteratively varied
so as to minimize the action. The method has been  criticised since the
galaxies are treated as point particles throughout their history, even
though the size of the galaxies must be comparable to their separation
at early times. However, the least action principle leaves the final
velocities of the galaxies unconstrained,
and its ability to reproduce the observed radial velocities
remains a  powerful  test of the validity of the trajectories. For the
Local  Group galaxies, Peebles has obtained remarkable agreement between the
observed radial velocities and those calculated from the least action
principle.

Supplied with a possible history for all members of the Local Group we
are able to evaluate the tidal torques acting on a particular galaxy at any
point along its orbit. Using a suitable model for the evolution of the
protogalaxy we can integrate the torque and investigate the acquisition of
the galaxy's spin angular momentum.  The magnitude of the angular momentum will
strongly depend on the details on the model.  However its direction
should not be as sensitive. Thus we will primarily restrict our results to
explaining the direction of the spin axis of the Milky Way and
Andromeda.

Gott and Thuan (1978) have studied the angular momentum of members of the Local
Group assuming it was isolated.  They observed that M31 and the Galaxy lie
nearly
in each other's plane and that their spins are opposite as measured along the
line of sight joining the two galaxies.  This strongly suggests that the spin
arose through tidal interaction.  Assuming that M31 and the Galaxy are
effectively isolated, they deduced that the original line of sight connecting
the two galaxies must be perpendicular to the spin angular momentum vectors
of both M31 and the Galaxy. This gives two possible trajectories.  We will
compare them with the ones given by the least action principle of Peebles (1989
and 1990). We will also test the hypothesis
that these two galaxies could be thought as isolated.  We will compare our
results with those of Raychaudhury and Lynden-Bell (1989) who have studied the
influence of external galaxies on the timing argument of Khan and Woltjer and
proposed possible orbits for  M31 and the Galaxy.

In the next section, we briefly review the least action method and determine
a set of trajectories for galaxies in the Local Group and dominant galaxies
in the Local Neighbourhood. In section~3 we calculate the orbital angular
momentum of the Local Group and compare this with the results from Raychaudhury
and Lynden-Bell(1989) as well as its implications for the analysis by Gott and
Thuan (1978). We present the tidal interaction picture in section~4 and
describe our adopted model of galaxy evolution. Finally, our results
are presented in section~5 and we draw conclusions in section~6.

\beginsec 2. Trajectories for Local Group galaxies.

In order to determine a set of trajectories for Local Group galaxies, we
follow the least action principle described by Peebles (1989 \&
1990, hereafter P1 and P2). However, in order to study the torque acting
on galaxies in the Local Group, we must also consider the effect
of the dominant galaxies and groups in the Local Neighbourhood.

\beginsubsec 2a. The Least Action Principle.

Peebles' least action principle  selects a set of classical
trajectories for a group of galaxies (point masses) which are interacting
through gravity, against
the background of an expanding universe model. This method differs from the
usual application of the least action principle in that boundary conditions
are applied to the beginning and end of each trajectory. The trajectories
are constrained such that
$$
\delta {\bf x}_i = 0 \ \ {\rm at} \ \ t = t_0, \hskip 1cm
a^2 d{\bf x}_i /dt \rightarrow 0 \ \ {\rm at} \ \ a\rightarrow 0
\eqno(2.1)
$$
where $a$ is the scale factor of the universe and ${\bf x}_i(a)$ is the
trajectory of the $i$th galaxy in comoving coordinates.
That is, the galaxies are fixed at their present positions at the present
epoch,
 and their peculiar velocities vanish as we approach the Big Bang.
Trial trajectories for the set of galaxies are adjusted in order to find
a stationary point in the action.

A $k=0$ universe with a possible cosmological constant is assumed, thus
$$
 H dt = { a^{1/2} da \over F^{1/2}},
\eqno(2.2)
$$
where $F=(\Omega + (1-\Omega)a^3)$, $H$ is the present Hubble constant and
$\Omega$ is the density parameter.
Following Peebles P2, the action for particles moving in such a universe
is,
$$
S= \int_0^{t_o} \Big [ \sum {m_i a^2 \over 2}
                \big ( {d{\bf x}_i \over dt} \big )^2
           + {G\over a} \sum_{i\neq j}{ m_im_j \over |{\bf x}_i - {\bf x}_j|}
           + {2\over 3} \pi G \rho_b a^2 \sum m_i{\bf x}_i^2 \Big ]
\eqno(2.3)
$$
{}from which we can deduce the equation of motion,
$$
a^{1/2}{d\over da} a^{3/2} {d \xb_i \over da}
 + {3(1-\Omega)a^4 \over 2F}{d\xb_i\over da}
  = {\Omega\over 2F} [ \xb_i + {R^3_0\over M_T}
  \sum_j {m_j(\xb_j-\xb_i)\over |\xb_j-\xb_i|^3} ].
\eqno(2.4)
$$
Here $R_0$ is the radius of a sphere which would enclose a homogeneous
distribution of the total mass $M_T$ of the group of galaxies considered,
$R_0 \equiv M_T ({4 \over 3} \pi \rho^0_b)^{-1}$.

It is very hard to have exact analytic solutions for the coupled
system of equations (2.4).  However Peebles succeeded in obtaining approximate
solutions using trial functions of the form
$$
\xb_i(a) = \xb^o_i + \sum_n {\bf C}^n_i f_n(a)
\eqno(2.5)
$$
where $\xb^o_i$ are the present positions of the galaxies and
the $f_n$ are a restricted set of `Fourier modes' chosen to satisfy the
boundary conditions (2.1).  The classical solutions are
obtained by introducing ${\bf x}_i(a)$ in the action and iteratively modifying
the
coefficient  ${\bf C}^n_i$ to obtain a stationary action.  In this paper we
take  $f_n = a^n (1-a)$ for $n=0, \ldots ,4$.  As Peebles did, we verify that
the least action solutions are good approximations to real solutions by
evolving the classical equations of motion starting with the initial positions
and velocities  derived from the least action solutions at $z=60$.

\beginsubsec 2b. Trajectories of Dominant Quadrupole Galaxies.

Peebles determined least action solutions for eight members of the Local
Group in P1 and also studied the influence from the Maffei and Sculptor groups
in P2. Here we will use the same Local Group members but also include five
galaxies/groups in the local neighbourhood which contribute a significant
torque on the Local Group. We adopt Peebles preferred universe from P2,
$H = 75$ km s$^{-1}$ Mpc$^{-1}$ and $\Omega = 0.1$. This will permit us to
study the influence of galaxies external to the Local Group in the past and
examine their effect on the spin of the galaxies.


Raychaudhury and Lynden-Bell (1989, hereafter RL) studied the effect
of nearby galaxies on the quadrupolar component of the gravitational
field at the barycentre of the Local Group. They showed that the galaxies
with the largest tidal influence on the Local Group lie within a radius of
7 Mpc. We follow their estimates for the distances to the most important
nearby galaxies, but represent the Sculptor group (NGC 55, 253, 247, 7793)
as a single point. Unlike RL, we also include the
Maffei galaxies at a distance of 3.6 Mpc despite their low galactic latitude
(b =$-1^\circ$) and the lack of any independent distance estimates. Buta and
McCall (1983) estimate a total extinction of at least five magnitudes to
Maffei~I.  However, there remains considerable uncertainty
over the distance to the Maffei group, and its corresponding mass.
Similar concerns surround the distance to IC~342 except that internal
absorption
is the main problem in this case. Freedman and Madore (1992)
estimate a total extinction of at least 2.2 magnitudes to IC~342.

Our final set of 13 galaxies is shown in table~1. Once again, we take the
lead from Peebles, P1, when determining galaxy masses and set the masses to
simplified ratios of the mass of M31. The Milky Way is taken
{}to be 70\% of mass of M31 whilst the others galaxies have relative masses
shown in table~1 determined from a comparison of their
corrected absolute blue magnitudes, based on the assumption that mass
traces light. The total mass is fixed by adjusting the mass of M31 such that
the least action trajectories reproduce the observed radial
velocity of M31, $v_{M31} = -123$ km s$^{-1}$. The resulting mass of M31 is
$M_{M31} = 10.8 \times 10^{12} M_\odot$ which corresponds to a
mass to light ratio of $M/L_B \approx 240$.

\beginsubsec 2c. Comparison of the Trajectories.

After minimizing to find a stationary point in the action, we obtain the
trajectories shown in figure~1.
The parameters used to obtain trajectories of figure~1 give a universe of 17
billions years.  In comoving coordinate, we can see that M31 and
the MW start from a distance between each other comparable with the distance
{}to galaxies outside the Local Group. Thus in the initial state of the
universe
the gravitational force on MW or M31 is dominated by galaxies outside the Local
Group.  This is true until the scale factor of the universe $a$ was half the
present value.  Therefore, before this epoch, it is not possible to regard the
effect of external galaxies as perturbations on the motion of Local Group
members.

{}From the LAP trajectories we can also deduce the maximum distance between M31
and the MW  which was $2.4 Mpc$ at  $a \sim 0.4$.  The mass of M31 has been
adjusted so that the projection of the velocity on the line of sight is $-123
km/s$.  However the LAP trajectories
permit us to calculate the other components of a galaxy's velocity. We find
a velocity for M31 of $436 km/s$ in the direction ($l=226^\circ$, $b=8^\circ$).
The proper motion of M31 is 0.6~km~s$^{-1}$~kpc$^{-1}$ ($1.26 \times
10^{-4}$~arcsec~yr$^{-1}$), which is twice the value of the prediction
made by RL, but still below observational limits. The components on the sky are
$\dot l = 1.26 \times 10^{-4}$~arcsec~yr$^{-1}$ and $\dot b = 4.8 \times
10^{-6}$~arcsec~yr$^{-1}$.
This perpendicular component of the velocity gives rise to a large angular
momentum for the Local Group.  Since the proper motion of M31 is more
significant than its line of sight velocity, we need a larger mass for M31 and
MW than is deduced using in the timing argument of Kahn and Woltjer (1959).

Comparing the trajectories of figure~1 with those obtained by Peebles,
we can see that adding more galaxies outside the Local Group leads to an
increase in the initial comoving separation of the Milky Way and Andromeda.
Since the external galaxies tend to pull the Milky Way and Andromeda apart, we
find that we need to increase the total mass of the system in order to
obtain a radial velocity of -123~km~s$^{-1}$ for Andromeda. A more massive
system has a larger value of $R_0$, which explains the increased separation.
We have also noted that the addition of these external galaxies degrades the
agreement for the line of sight velocity of Local Group members
that Peebles had obtained for $\Omega=0.1$ and $H=75$. In our case a higher
value of $\Omega$ may give  better results.  However this does not change the
qualitative behaviour of the LAP trajectories, we will therefore
retain Peeble's values of $H$ and $\Omega$.

\beginsec 3. Orbital Angular Momentum in the Local Group.

Gott and Thuan (1978) have attempted to calculate the relative initial position
of the Milky Way and Andromeda assuming that they form an isolated system
which conserves angular momentum (spin and  orbital).  They also assumed,
as we do, that the spin of these galaxies would arise from mutual tidal
interaction.  From these two premises they concluded that the original
direction between these galaxies is perpendicular to the direction of their
spins.  This gave the result that the   original direction of Andromeda with
respect to the Milky Way must have been  either $l=152^\circ, b=0^\circ$ or
$l=332^\circ, b=0^\circ$.

However, if we examine the relative velocity of M31 and the Milky Way using
our trajectories, the magnitude of the angular momentum of the pair with
respect to their common centre of mass is $8 \times 10^{77} g\, cm^2/s$. This
is much larger than their individual spin angular momenta which is of the
order of $10^{74} g\, cm^2/s$.  Thus the least action principle implies
that the Local Group can not be thought of as a tidally isolated system.

Recently, Raychaudhury and Lynden-Bell (1989) showed that there is an
appreciable quadrupole moment acting upon the Local Group even at the present
epoch. Assuming that the eigendirections of the quadrupole moment did not
substantially change direction and that the tidal force was only a small
perturbation
they calculated possible trajectories for Andromeda using the equation
$$
\ddot {r^\alpha} - GM_t{{r^\alpha} \over |{\bf r}|^3}
      = \sum_\beta Q^{\alpha\beta}r^\beta .
\eqno (3.1)
$$
${\bf r}$ is the separation vector between the Milky Way and Andromeda,
$M_t$ their total mass and $\sum_\beta Q^{\alpha\beta}r^\beta$ is the
projection of the quadrupole component of the gravitational field in
the radial direction. This equation assumes that the tidal force
$Q\cdot r$ is weak compared to the inverse square law.  They obtained
various trajectories by assuming different time dependencies of the
eigenvalues of the  quadrupole moment.  However, using the trajectories
{}from section~2, we can calculate the quadrupole moment acting upon the
Local Group as a function of the scale size of the universe. These are
presented for $a=0.1$, $a=0.5$ and $a=1.0$ in figure~2 where the
projection of the  quadrupole on the line of sight is illustrated.
Table~2 summarizes the eigenvalues and eigendirections of
$Q^{\alpha\beta}$ in each case. We can see that the eigendirections
have changed since the Big Bang and that might have affected the tidal
force.  In figure~3 we present the trajectory of Andromeda obtained by
Raychaudhury and Lynden-Bell and compare it with the one from the least
action principle.  They are qualitatively very similar.  However our
trajectory for $M31$ is longer.  This is  partly a result of using
different values for $\Omega$ and $h$ but also due to the fact that the
tidal force due to local neighbourhood galaxies is not a perturbation
in the very early universe and thus the equation used by Raychaudhury
and Lynden-Bell is invalid.


\beginsec 4. Quadrupole---quadrupole Interaction.

Hoyle (1949) has suggested that the spin of a galaxy might result from the
tidal
interaction with neighbouring galaxies. Supplied with our set of trajectories
for galaxies in the local neighbourhood, we can concentrate on any individual
galaxy and determine the tidal torque acting upon it throughout
its history. Using a model, which represents the selected galaxy as a rigid
spheroid, we use the quadrupole-quadrupole interaction to numerically
integrate the tidal torque acting on the selected galaxy and determine its
final
spin angular momentum.

The change in spin angular momentum is given by the torque
$$
\Gamma_\alpha = {d I_{\alpha\beta}\omega_\beta \over dt} .
\eqno(4.1)
$$
where $I_{\alpha\beta}$ is the moment of inertia tensor of the selected galaxy,
${\bf \omega}$ its frequency of rotation.

The torque due to the gravitational field generated by the
distribution of point mass galaxies surrounding the selected galaxy is,
$$
\Gamma_\alpha = {1\over 3} \sum_{\beta ,\gamma} \epsilon_{\alpha\beta\gamma}
                q_{\beta\delta} Q_{\delta\gamma}
\eqno (4.2)
$$
where
$$
 Q_{\delta\gamma} = \sum_{i} { GM_i \over |\xb -\xb_i|^3}
           \Big[ {3(\xb -\xb_i)_\delta(\xb -\xb_i)_\gamma
                                 \over |\xb-\xb_i|^2} - \delta_{\delta\gamma}
           \Big]
\eqno (4.3)
$$
is the quadrupole component of the gravitational field and $q_{\beta\delta}$ is
the quadrupole moment of the selected galaxy.

\beginsubsec 4a. The Galaxy Model.

In order to calculate the spin of the selected galaxy, we have to know its
inertia and
quadrupole moments. Since the dark halo is the dominant component of a galaxy
for gravitational interactions, we  model the galaxies as
spheroids. In fact in this paper the galaxy spheroids are rigid, which
can only be approximation since in the absence of  dissipation,
protogalaxies which acquire their spin through
gravitational interactions must have zero vorticity as no vorticity is observed
in the microwave background.
However, tidal forces are linear with distance
{}from the center of mass and thus induce a state of approximate solid body
rotation (Gott 1975), so this seems to be a reasonable approximation.

The quadrupole moment of the selected galaxy is,
$$ q_{ij} ={m \over5}  R^2(1-\epsilon^2) diag (1,1,-2)
\eqno (4.4)
$$
in a frame defined by the axis of symmetry of the spheroid.
The five undetermined parameters are the mass $m$ of the galaxy,
the length of the semi-major axis $R$, the eccentricity $\epsilon$ of the
ellipse of revolution and two angles specifying the direction of the
major axis.

In the gravitational instability picture small density perturbation of
sufficient
magnitude expand to a maximum size $R_m$ and recollapse.  In an $\Omega=1$
universe the radius of the proto-galaxy obeys the equation
$$
R = {R_m\over 2} (1 - \cos \eta),
$$
$$
a = {a_{\rm col}\over (\pi)^{3/2}} (\eta - \sin\eta)^{2/3}
\eqno (4.5)
$$
where $a_{\rm col}$ is the scale factor of the universe when the galaxy
starts to recollapse.  There is the following relation between $R_m$
and $a_{\rm col}$
$$
a_{\rm col}\approx \Big ({H^2\pi^2\over 2 GM}\Big )^{1/3} R_m .
\eqno(4.6)
$$
$R_m$ remains a free parameter and is related to the time at which galaxies
form. Binney and Silk have also shown that the  eccentricity initially varies
proportional to the scale factor
of the universe, but rapidly becomes a quadratic variation. Here we
will simply model the evolution of the eccentricity of a galaxy spheroid by,
$$
\eqalign{
\epsilon &=  1-\epsilon_m(a/a_{\rm col})^2  \hskip 1cm a<a_{\rm col} \cr
         &=  1-\epsilon_m                \hskip 2.35cm a>a_{\rm col} \cr}
\eqno(4.7)
$$
We will present results for the case of $ \epsilon_m = 0.81$ but we have also
examined other values, all of which give a qualitatively similar picture.
A more elongated spheroid has a   stronger coupling to the tidal field,
resulting
in a larger spin angular momentum. Since the eccentricity of the galactic
halo is essentially unknown, we do not feel that we can accurately predict
the magnitude of the spin of any particular galaxy, but the direction of
the spin axis is less sensitive to the shape of the galaxy model and
so provides a better indicator.

\beginsubsec 4b. Initial Orientation of a Proto-galaxy.

The remaining free parameter of the model is the initial orientation of
the major axis of the selected galaxy.  We find that the orientation
of the final spin axis is very sensitive to this parameter.
This can be seen in figure~4 which plots the direction of the final spin
axis calculated for the Milky Way and Andromeda as a function of the
orientation of their major axes at $a=0.1$. Successive evaluations were made
for initial orientations based on a grid spaced by $10^\circ$ in $l$ and $b$,
assuming $a_{\rm col} = 0.5$. We plot
the  projection of the unit spin vector of the model onto the unit vector of
the observed spin. The lightest contour level corresponds to an
angle, $\theta \leq 15^\circ$, between the two vectors. The contours increase
by  $15^\circ$ intervals up to $\theta = 90^\circ$, with
the final contour at $\theta = -45^\circ$. The symmetry comes from the
bi-polar symmetry of the galaxy spheroid model.

By definition the direction of the observed spin for the Milky Way is,
$$
\hat {\bf J}_{MW}: b = -90^\circ,
$$
whilst for Andromeda we adopt,
$$
\hat {\bf J}_{M31}: l = 242^\circ , b = -30^\circ
$$
quoted by Gott and Thuan. However, these spin axes are determined from
the rotation of the luminous matter in the galaxy, whereas our model
more accurately represents the the dominant mass and therefore the dark
halo component of the galaxies. Kuijken (1991) modeled the disc warps
resulting from the misalignment between the luminous disc and the halo
of a spiral galaxy. The Galaxy is observed to have a moderate disc
warp, and he estimates that there could be a discrepancy as large as
$30^\circ$ between the observed spin axis and the actual spin of the
halo.

{}From figure 4a we can see that a random choice for the initial orientation
of the Milky Way would have approximately a
 1.7\% chance of producing a final spin
within $15^\circ$ of the observed direction. The a priori chance of selecting
a vector within $15^\circ$ of another is 3.4\%. Conversely, the similar
probability for Andromeda is $\sim$ 5.8\%.

However, we do not have an entirely free hand when selecting the initial
orientation for a galaxy.
Binney and Silk (1979) have shown that a first
order effect of tidal interaction is to produce an anisotropy in the density
perturbation of a   protogalaxy.  The anisotropy is aligned so as to minimise
the torque on the galaxy.  This inhibits
the early rotation of the galaxy, but gives a preferred direction for the
initial orientation of its major axis. As the matter distribution around the
galaxy changes, there will be a change in the direction of the local axis of
the tidal field which will induce the galaxy to spin. In
figure~5 we have plotted the magnitude of the tidal field $\propto m/d^3$ for
the Galaxy and Andromeda as a function of the scale size of the universe.
At the present epoch, Andromeda provides the dominant contribution to the
quadrupole acting on the Milky Way and vice versa.  However, at early times
the quadrupole moment acting on them both is dominated by galaxies
external to the Local Group.  In both cases the
transition in the quadrupole occurs at a redshift of between $z=0.5$ and $z=1$.
{}From qualitative arguments we see that if these galaxies collapse much
before this transition time ($a_{\rm col} < 0.4$), very little angular momentum
will be transferred to them.

\beginsec 5. Results.

So far we have not fixed the
value of $a_{\rm col}$. At the time of writing, there is little observational
evidence for a population of proto-galaxies at $z<1$, which argues against
$a_{\rm col}$ being much larger than $0.5$.
On the other hand, we find that it is   comparatively difficult to get
the galaxy to spin at all if $a_{\rm col} < 0.3$.
This is easy to see qualitatively. With the Milky Way pointing towards
IC~342 initially, there is little change in the direction of the eigenvectors
of ${Q_{\delta\gamma}}$ until $a=0.3$. Then as the separation between the Milky
Way and Andromeda decreases, Andromeda comes to dominate the quadrupole moment
and the major axis of the Milky Way swings round to point towards Andromeda.
After $a_{\rm col}$, $q_{\rm MW}$ becomes small and the galaxy
effectively decouples from the tidal field, rotating faster as it collapses,
but with little change in the direction of the spin axis. Hence $a_{\rm col}$
should lie in the range 0.3 to 0.5. We have opted for $a_{\rm col} = 0.5$
since this results in a larger spin angular momentum.

The position of the galaxy which dominates the quadrupole moment at
$a=0.1$ provide a   preferred direction for the initial orientation of the
selected galaxy.  From figure~5 we can see that the  preferred choice for the
initial orientation of the Milky Way is towards IC~342 ($l = 109^\circ$, $b =
06^\circ$ at $a=0.1$). This gives
  disappointing results, with the final spin axis in the direction
($l=193^\circ$, $b=-12^\circ$) -- approximately $ 80^\circ$ away from the
observed direction.

We have investigated the possibility that initial quadrupole is not dominated
by IC342 but rather by N5128. This might be a possibility if the mass of
$N5128$ has been underestimated by a factor of a few percent or the distance
by a factor of $1.2^{1/3}$. Using NGC~5128 to define the initial alignment,
leaves the galaxy rotating with the spin axis oriented towards
($l= 327^\circ$, $b=-74^\circ$), within $16^\circ$ of the
observed direction.  The resulting angular momentum is $5.2\times 10^{74} g
cm^2/s$.
A similar situation exists with Andromeda. In this case the Maffei group
would appear to dominate the early quadrupole. Aligning Andromeda towards
it gives a spin in the direction of ($l=44.1$, $b=-0.6$). Pointing
towards N5128 is little better, but the M81 group yields the result ${\bf
J}_{M31}$:
($l=238^\circ$, $b=-38^\circ$). This is about $9^\circ$ away from the observed
direction, so once again we are in remarkably good agreement
with the observations.  In this case the obtained magnitude is
$1.0\times 10^{75} g~cm^2/s$.

For completeness, figure~4a shows the direction from the Milky Way to
the other galaxies in the sample at $a=0.1$ and figure~4b shows the
positions relative to Andromeda. In both cases, only one galaxy lies
within the top contour which corresponds to a final spin less than
$15^\circ$ from the observed direction.

\beginsec 6. Conclusion.

We have investigated the angular momentum of the Local Group and its largest
galaxies, the Milky Way and Andromeda, using the least action principle of
Peebles.  We have augmented the sample of galaxies to
include galaxies from the Local Neighbourhood which contribute significantly to
the present torque.  We have  found that the Local Group has a large angular
momentum
of order $8\times 10^{77} g cm^2/s$ in the direction
($l=75^\circ$,$b=75^\circ$).
This implies that the Local Group cannot be thought of as tidally isolated,
and the total angular momentum in the Local Group is not conserved.

We have also investigated the angular momentum of the Milky Way and Andromeda.
We have found that there is a transition time ($z\approx 1$) when  the tidal
forces on them
is dominated by Local Neighbourhood galaxies and thereafter Local Group ones.
The least action trajectories were able to reproduce the direction of the  spin
of these galaxies within $15^\circ$.
In principle, once the trajectories of the members of the Local Group are
known,
we are in a position to integrate the torque acting on an individual galaxy
and determine the final amplitude of the angular momentum. This might provide
another
test of the least action trajectories, but in practice we find that the
results are more sensitive to the details of the model used to represent
the evolution of the protogalaxy.
\bigskip

\vskip 0.25 truein

\noindent{\bf Acknowledgments}.  We would like to thank D. Lynden-Bell for
useful comments and S. Raychaudhury who provided us with a his catalogue
of nearby galaxies derived from the Kraan-Korteweg catalogue.

\vfill \eject
\openup 1\jot

\beginsec 7. References

\noindent
Barnes, J. \& Efstathiou, G., 1987, {\it Astrophys.J.}, {\bf 319}, 575.

\noindent
Binney, J. \& Silk, J., 1979, {\it M.N.R.A.S.}, {\bf 188}, 273.

\noindent
Buta, RJ. \& McCall, M.L., 1983, {\it M.N.R.A.S.}, {\bf 205},131.

\noindent
Gott, J.R., 1975, {\it Astrophys.J.}, {\bf 201}, 296.

\noindent
Gott, J.R. \& Thuan, T.X., 1978, {\it Astrophys.J.}, {\bf 223}, 426.

\noindent
Hoyle, F., 1949, In {\it Problems of Cosmic Aerodynamics}, (International
Astronomical Union).

\noindent
Kahn, F.D. \& Woltjer, L., 1959, {\it Astrophys.J.}, {\bf 130}, 705.

\noindent
Kuijken, K., 1991, {\it Astrophys.J.}, {\bf 376}, 467.

\noindent
Madore, B.F. \&  Freedman, W.L., 1992 PASP {\bf 104}, 362


\noindent
Peebles P.J.E., 1969, {\it Astrophys.J.}, {\bf 155}, 393.


\noindent
Peebles, P.J.E., 1989, {\it Astrophys.J.}, {\bf 344},L53.

\noindent
Peebles, P.J.E., 1990, {\it Astrophys.J.}, {\bf 362}, 1.

\noindent
Rychaudhury, S. \& Lynden-Bell, D., 1989, {\it M.N.R.A.S.}, {\bf 240}, 195.

\noindent
White, S., 1984, {\it Astrophys.J.}, {\bf 286}, 38.

\openup -1\jot
\vfill \eject
\bigskip

\vbox{\tabskip=0pt \offinterlineskip
\halign to\hsize{\strut#&#\tabskip=1em plus2em minus.5em&
#\hfil &#& \hfil# &#& \hfil# &#& \hfil# &# & \hfil# &#&
\hfil# &#& \hfil# &#& \hfil# &#& #\hfil &\hfil# \cr

&&  \omit\hidewidth Name \hidewidth &&
$l $&& $b $&&
\omit\hidewidth  $R$  \hidewidth &&
\omit\hidewidth $v_{CLG}$ \hidewidth &&
\omit\hidewidth $M_r$ \hidewidth &&
\omit\hidewidth  $m_B^{0,i}$ \hidewidth &&
\omit\hidewidth type \hidewidth &\cr
&&   &&   &&   && \omit\hidewidth (Mpc) \hidewidth &&
\omit\hidewidth (km~s$^{-1}$) \hidewidth &&        &&   &&   &\cr
&&           &&       &&       &&      &&         &&        &&        &&
&\cr
&& {\it local group}  &&       &&       &&       &&         &&        &&
&&    &\cr
&&Milky Way  &&  ---  &&  ---  &&  0.0 &&   0.0   &&  0.700 &&    --- && S
&\cr
&&NGC~6822   &&  25.3 && -18.4 &&  0.5 && -179.8  &&  0.025 &&   9.35 && I
&\cr
&&M~31       && 121.2 && -21.6 &&  0.7 && -539.7  &&  1.000 &&   4.38 && S
&\cr
&&IC~1613    && 129.7 && -60.6 &&  0.7 && -348.3  &&  0.025 &&   9.99 && I
&\cr
&&WLM        &&  75.9 && -73.7 &&  0.9 && -198.0  &&  0.025 &&  11.04 && IB
&\cr
&&Sextans~AB && 246.2 &&  39.9 &&  1.3 &&  113.4  &&  0.025 &&  11.87 && IB
&\cr
&&NGC~3109   && 262.1 &&  23.1 &&  1.7 &&  129.6  &&  0.100 &&  10.39 && S
&\cr
&&NGC~300    && 299.2 && -79.4 &&  2.2 &&   96.9  &&  0.100 &&   8.70 && S
&\cr
&&           &&       &&       &&      &&         &&        &&        &&
&\cr
&& {\it local neighbourhood} \span \span       &&       &&       &&         &&
      &&        &&    &\cr
&&Sculptor   && 105.8 &&  85.8 &&  3.2 &&  221.1  &&  2.000 &&   9.00 && S
&\cr
&&Maffei~AB  && 136.4 &&  -0.4 &&  3.5 &&  199.0  &&  2.000 &&  14.80 &&
E/S&\cr
&&M~81       && 142.1 &&  40.9 &&  3.5 &&  103.3  &&  0.410 &&   7.86 && S
&\cr
&&NGC~5128   && 309.5 &&  19.4 &&  4.9 &&  307.8  &&  1.500 &&   7.89 && L
&\cr
&&IC~342     && 138.2 &&  10.6 &&  6.1 &&  227.7  &&  4.000 &&   9.42 && S
&\cr
} }
\bigskip
\noindent{\narrower\smallskip
  \noindent{\bf Table 1.} The sample of nearby galaxies and their properties.
The galaxies are divided into two sections, (i) Local Group galaxies studied
by Peebles and (ii) other dominant galaxies up 7 Mpc from the Local Group.
The columns are (a) Name, (b,c) present galactic coordinates, (d) distance from
the galaxy, (e) present velocity with respect to the centre of the Local Group,
(f) Mass relative to the mass of M31, (g) Corrected apparent blue magnitude,
and (h) Hubble type.
\smallskip}

\bigskip
\vfill \eject
\bigskip

\vbox{\tabskip=0pt \offinterlineskip
\halign to\hsize{\strut#&#\tabskip=1em plus2em minus.5em&
#\hfil &#& \hfil# &#& \hfil# &#& \hfil# &\hfil# \cr

&&  \omit\hidewidth Scale Factor \hidewidth &&
\omit\hidewidth  $Eigenvalues$  \hidewidth &&
$l\, $&& $b $ &\cr
&&   &&   &&   &&    &\cr
&&   &&   &&   &&    &\cr
&& a=0.1  &&  313 && 302   &&  22  &\cr
&&        && -240 && 213   && -0   &\cr
&&        &&  73.2 && 303   && -67  &\cr
&&   &&   &&   &&    &\cr
&& a=0.5  &&  3.24 && 126   && -17  &\cr
&&        &&  0.96 && 317   && -71  &\cr
&&        && -4.20 &&  37   &&   3  &\cr
&&   &&   &&   &&    &\cr
&& a=1.0  &&  1.45 && 138   &&  33  &\cr
&&        &&  0.29 && 132   && -56  &\cr
&&        && -1.75 && 227   && -3  &\cr
}}
\bigskip
\noindent{\narrower\smallskip\noindent{\bf Table 2.} Eigenvalues and
eigendirections of $Q^{\alpha\beta}$ evaluated at the barycentre
of the Local Group for three different scale factors $a$.  This shows the
importance of the variation of the eigendirections of the quadrupole moment as
a function of time.  The eigenvalues are in units of
$0.43\times 10^{12}{\rm M}_\odot/{\rm Mpc}^{3}$.\smallskip}

\bigskip
\vfill \eject
\proclaim Figure captions.

\vskip 0.25 truein
\openup -1\jot
\noindent {\narrower\smallskip\noindent Figure 1.  Least action trajectories
for galaxies of Table 1.  The $x$ direction is towards Andromeda at $\alpha=
10^\circ\ ,\delta=41^\circ $, the $y$ direction towards $\alpha= 100\ ,\delta=
0^\circ$ and the $z$ direction $\alpha=190^\circ \ ,\delta=  49^\circ$.
\smallskip}     \openup 1\jot  \vskip 0.25 truein

\vskip 0.25 truein
\openup -1\jot
\noindent {\narrower\smallskip\noindent Figure 2. History of the tidal force
$Q . r$ due to galaxies external to the Local Group, projected on the sky
viewed from the center of mass of the Local Group.  The three projections
correspond to $a=0.1,\ 0.5,\ 1.0$ showing that the eigendirections have changed
 since the Big Bang. The eigenvalues and eigendirections are also shown in
table~2.
\smallskip}
\openup 1\jot  \vskip 0.25 truein

\vskip 0.25 truein
\openup -1\jot
\noindent {\narrower\smallskip\noindent Figure 3.  The sky projection of
the least action trajectory for M31 from this paper (the plain line)
compared with the trajectory
obtained by Raychaudhury and Lynden-Bell (the dotted line).  The least action
trajectory is slightly longer, partly due to a different choice of
the density parameter $\Omega$ and Hubble constant $H$. It is also due to the
fact that for the least action trajectories, it is not possible to assume that
the quadrupole force from galaxies outside the Local Group is only a
perturbation.
In the early universe these later galaxies dominate the force on the Milky
Way and Andromeda.
\smallskip}
\openup 1\jot  \vskip 0.25 truein

\vskip 0.25 truein
\openup -1\jot
\noindent {\narrower\smallskip\noindent Figure 4a.  The projection of the
unit spin of the Milky Way onto the unit vector in the direction of the
observed spin as a function of l,b of the orientation of the galaxy model.
\smallskip}
\noindent {\narrower\smallskip\noindent Figure 4b. The corresponding result
for Andromeda.
\smallskip}
\openup 1\jot  \vskip 0.25 truein

\vskip 0.25 truein
\openup -1\jot
\noindent {\narrower\smallskip\noindent Figure 5. Quadrupole moment on the
Galaxy and M31 in function of the radius of the universe and $z$.  There is a
transition time ($a\approx 0.5-0.7$) during which the dominant contribution
{}to the quadrupole moment passes from galaxies outside the Local Group to
those within.

\smallskip}
\openup 1\jot  \vskip 0.25 truein

\bye